\documentclass[showpacs,aps,prb,twocolumn,floatfix]{revtex4}
\usepackage{amssymb,amsfonts}
\usepackage{times}
\usepackage{graphicx}
\usepackage{bm}
\usepackage{amsmath}

\newcommand{\il}{\int\limits}
\newcommand{\be}{\begin{equation}}
\newcommand{\e}{\end{equation}}
\newcommand{\beml}{\begin{subequations}}
\newcommand{\eml}{\end{subequations}}
\newcommand{\beq}{\begin{eqnarray}}
\newcommand{\eq}{\end{eqnarray}}
\newcommand{\ba}{\begin{array}}
\newcommand{\ea}{\end{array}}
\newcommand{\lt}{\left}
\newcommand{\rt}{\right}
\newcommand{\n}{\nonumber}

\newcommand{\ep}{\varepsilon}

\newcommand{\bb}{\boldsymbol}

\DeclareMathOperator{\tr}{Tr}

\DeclareMathOperator{\im}{Im}
\DeclareMathOperator{\re}{Re}

\begin{document}
 
\date{August, 2008}

\title{Thermopower oscillations in mesoscopic Andreev interferometers}

\author{M.~Titov}

\affiliation{School of Engineering \& 
Physical Sciences, Heriot-Watt University, Edinburgh EH14 4AS, UK
}

\begin{abstract}
Anomalously large thermopower of mesoscopic normal-metal/superconductor interferometers
has been investigated by Chandrasekhar {\it et al.} 
It was shown that, depending on the geometry of the interferometer, 
the thermopower is either symmetric or antisymmetric
periodic function of the magnetic flux. We develop a detailed theory 
of the observed thermoelectric phenomena in the framework 
of the non-equilibrium quasiclassical approach. 
In particular we provide, for the first time, a possible explanation of the  
symmetric thermopower oscillations. This effect is attributed to
the electron-hole symmetry violation that originates in the steady-state charge 
imbalance between different arms of the interferometer. Our theory can be tested
by an additional control over the charge imbalance in a modified setup geometry.  
We also predict a sign reversal behavior of the thermopower with increasing
temperature that is consistent with the experiments by Parsons {\it et al.} 
\end{abstract}
\maketitle

\section{Introduction}
\label{sec:intro}

A linear response of a metal to the temperature gradient $\bb{\nabla} T$
and external electric field ${\bb \cal E}$ is characterized 
by four transport coefficients:
$\sigma, \kappa, \eta, \zeta$, which are defined, 
under the open circuit conditions, 
by the following relation\cite{Abrikosov} 
\be
\label{LR}
\lt(\ba{c} \bb{J}^e\\ \bb{J}^Q\ea\rt)=
\lt(\ba{cc} \sigma & \eta \\ \zeta & \kappa \ea\rt)
\lt(\ba{c} \bb{\cal E}\\ -\bb{\nabla} T \ea\rt),
\e
where $\bb{J}^e$ is the electric current and $\bb{J}^Q$ is the heat flow 
induced in the circuit. The diagonal terms, $\sigma$ and $\kappa$, 
are the electric and thermal conductivity, while the off-diagonal ones, 
$\eta$ and $\zeta$, are the thermoelectric coefficients that 
fulfill the Onsager relation $\zeta=T \eta$.  

The thermoelectric response is usually observed in a
confined geometry by measuring the voltage rather 
than the electric current. The ratio of the measured voltage to 
the temperature gradient applied across the sample 
is known as the Seebeck coefficient (or, the thermopower)
and is given by $S=\eta/ \sigma=(\bb{\cal E}/\bb{\nabla} T)|_{\bb{J}^e=0}$.
The Mott's formula\cite{Mott1936} relates the thermopower to 
the energy dependent conductivity $\sigma(\ep)$
\be
\label{S}
S=-
\frac{\pi^2}{3}\frac{k_B^2T}{e} 
\lt.\frac{d \ln \sigma(\ep)}{d\ep}\rt|_{\ep=\ep_F},
\e
and illustrates an important role of the electron-hole asymmetry
for the thermoelectric response in metals. Indeed, 
one can see from Eq.~(\ref{S}) that $S$ vanishes 
if $\sigma(\ep)$ is symmetric around the Fermi energy $\ep_F$. 
In the case of a spherical Fermi surface the thermopower
$S\sim k_B^2 T/ e\,\ep_F$ is finite only because of 
a tiny difference in the effective masses of electrons and holes.

In superconductors the situation is complicated by 
the presence of Bose condensate of Cooper pairs
that does not respond to a temperature gradient. 
It was suggested by Ginzburg\cite{Ginzburg1944}  
on the basis of the two-fluid model of superconductivity 
that the linear response theory (\ref{LR}) still applies to 
a normal (dissipative) component of the electron liquid. 
The thermoelectric effect is, however, shunted by 
the superconducting component and cannot be observed
in bulk superconductors.\cite{Steiner1935,Huebener1972}
An incomplete cancellation of the thermoelectric and supercurrents 
was, nevertheless, predicted for anisotropic superconductors
and superconducting bimetallic samples.\cite{Galperin1973,Garland1974}
This theory has been experimentally confirmed by 
Zavaritskii\cite{Zavaritskii1974} by using a bimetallic loop 
to detect the thermopower in a superconducting state. 
The method relies upon the quantization of the superconducting 
condensate in the loop that prevents a complete cancellation 
of the thermoelectric current and generates a small, but measurable, 
magnetic flux $\Phi_T\sim 10^{-2} \Phi_0$,
where $\Phi_0=2e/h$ is the flux quantum.
In subsequent experiments,\cite{Harlingen1980} however,
much greater values of the thermoelectrically induced 
flux, $\Phi_T\sim 10^2 \Phi_0$, were detected.
The origin of this "giant flux" puzzle 
remains debated.\cite{Kolachek2004,Gurevich2006}

\begin{figure}[t]
\includegraphics[width=0.98\columnwidth]{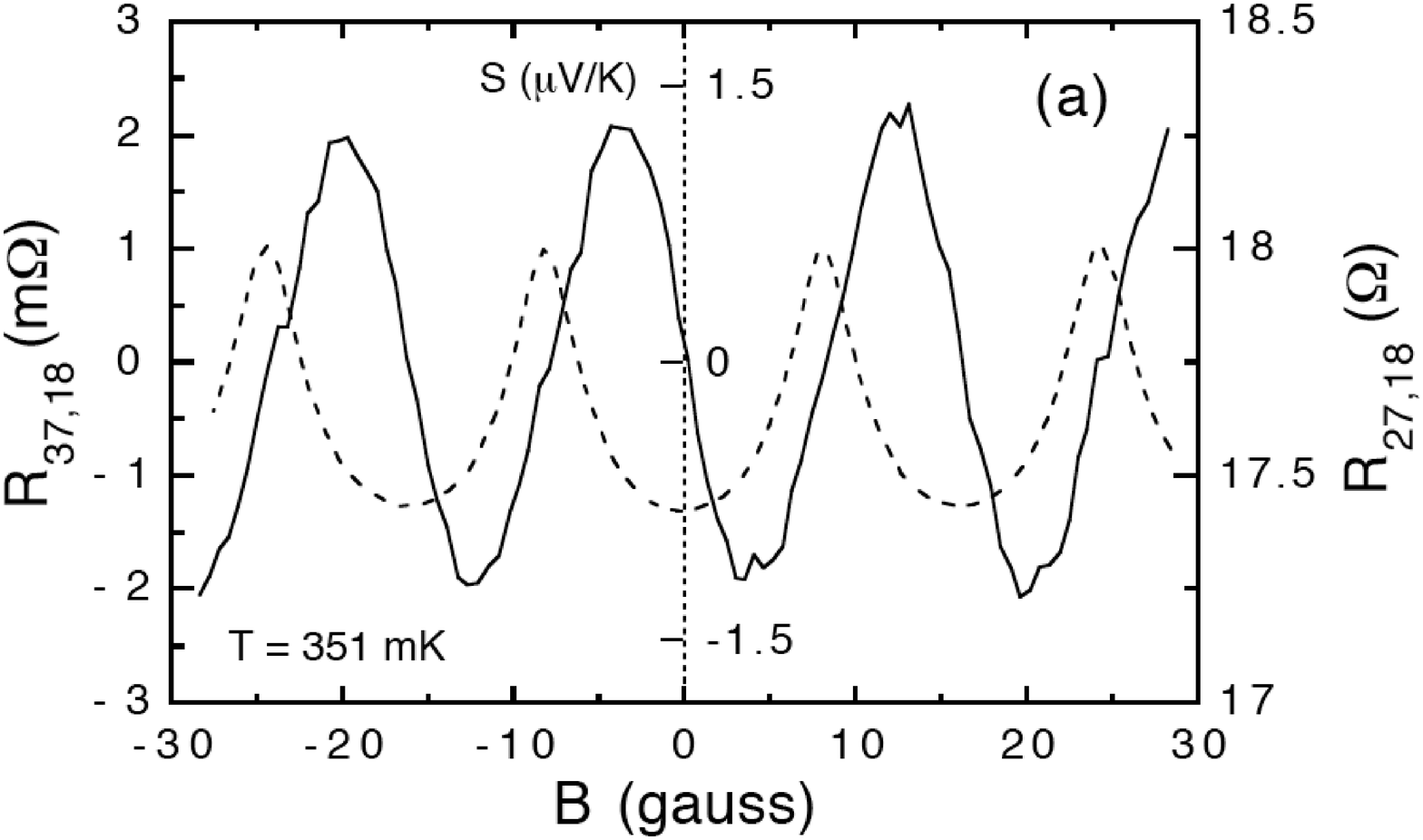}
\includegraphics[width=0.98\columnwidth]{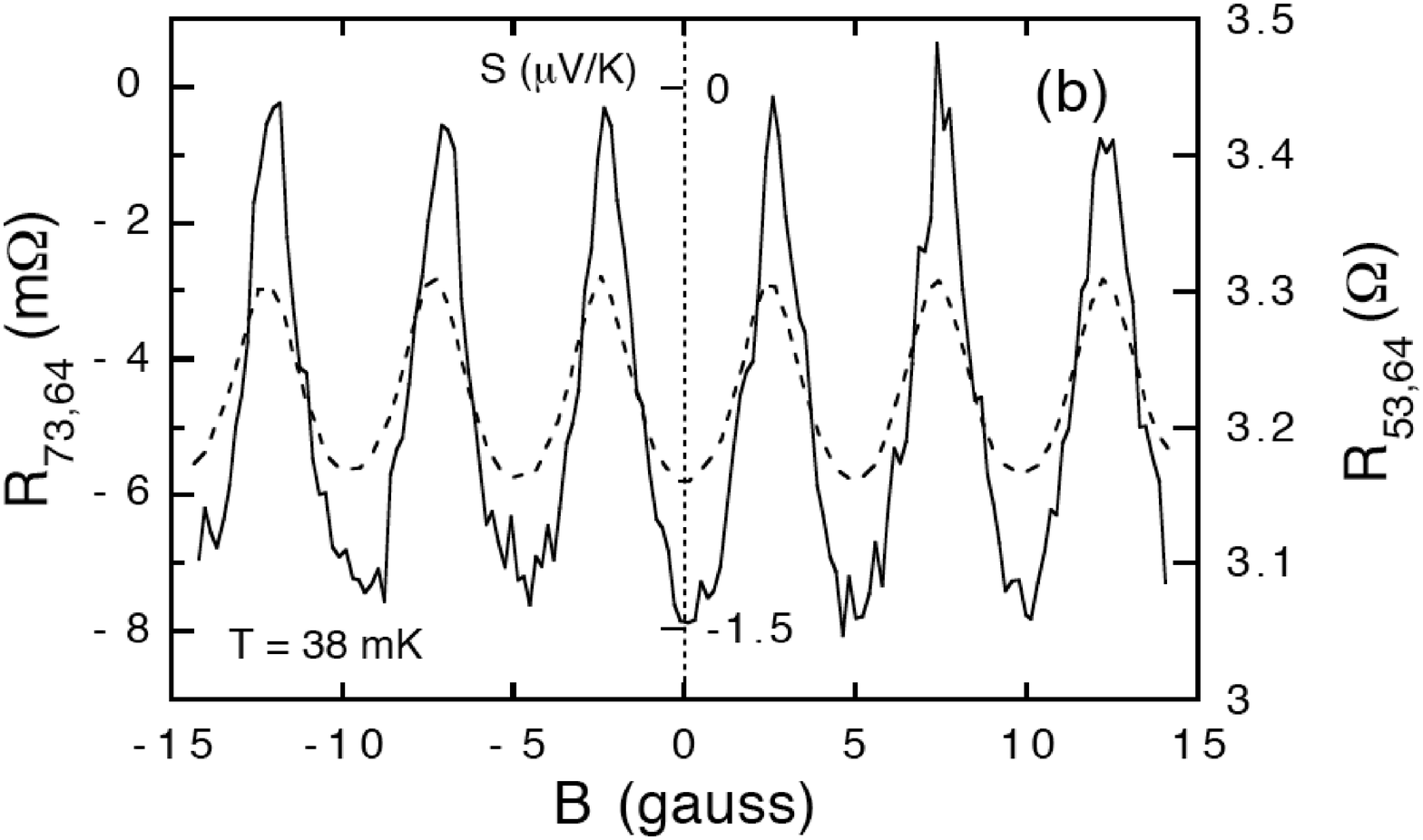}
\caption{
The experimentally observed thermopower oscillations (solid line) 
in the parallelogram (a) and in the house (b) interferometer.
The figure is adopted from Ref.~\onlinecite{Eom1998}.
}
\label{fig:exp}
\end{figure}

Mesoscopic systems provide an altogether different 
way to probe the thermoelectric phenomena in the 
presence of superconductivity. The superconducting 
correlations can penetrate the normal-metal part 
of the system due to a process known as Andreev reflection. 
As the result the thermopower of a mesoscopic
normal-metal wire can be affected by contacting the wire 
to a superconductor. First experiments of this kind 
have been performed a decade ago by Chandrasekhar 
{\it et al.}\cite{Eom1998} (see Fig.~\ref{fig:exp}). 
In these and in numerous subsequent 
experiments\cite{Eom1999,Eom2000,Dikin2002,
Parsons2003,Parsons2003b,Jiang2005a,Jiang2005b,Jiang2005c,Cadden2007}  
the thermopower of a normal-metal wire in proximity 
to a superconducor/normal-metal loop was found to oscillate
as a function of the magnetic flux piercing 
the loop. The amplitude of the oscillations 
was shown to exceed the thermopower of the normal-metal wire 
in the absence of proximity-induced superconductivity.
The experimental results suggest that the proximity effect
is responsible for the electron-hole symmetry violation in these systems, 
which is no longer suppressed by a small factor $k_B T/\ep_F$.
The lack of the electron-hole symmetry in the experiments 
was attributed long ago\cite{Seviour2000,Kogan2002} 
to a voltage induced between the superconducting 
and the normal-metal part of the mesoscopic circuit 
under non-equilibrium conditions. This voltage is directly related  
to the shift between the chemical potential in the normal metal
and that in the superconductor. Similar shift, which is commonly referred 
to as the charge imbalance, can exist inside a bulk 
superconductor between the chemical potential of quasiparticles and 
that of the Cooper pairs. 

The charge imbalance in bulk superconductors and its role in the thermoelectric effects
were intensely studied in 1970s (see Ref.~\onlinecite{Pethick1980} for the review.). 
The imbalance can be achieved either by injection of a dissipative current 
to the superconductor\cite{Rieger1971,Yu1972,Clarke1972,Tinkham1972} or by applying a temperature gradient 
in the presence of a superconducting flow\cite{Pethick1979,Clarke1979,Clarke1980} (Pethick-Smith effect).
The relaxation of the charge imbalance is determined primarily by the electron-phonon interaction.
From the detailed theory provided by Schmid and Sch\"on\cite{Schmid1979,Schmid1975} 
one can roughly estimate the imbalance relaxation time in bulk $s$-wave superconductors 
as $\tau_Q \sim (k_B T/\Delta) \tau_{in}$, where $\Delta$ is 
the value of the superconducting order parameter and $\tau_{in}$ 
is the temperature dependent electron-phonon scattering time.
In our study we deal instead with the voltage drop between 
the superconducting and the normal-metal arms of the interferometer. 
We, however, keep using the term imbalance to stress 
the analogy to the thermoelectric effects in bulk superconductors. 

The non-equilibrium quasiclassical theory of proximity effect has been successfully
applied in Refs.~\onlinecite{Virtanen2004a,Virtanen2004b,Virtanen2007,Chandrasekhar2004} 
to describe the antisymmetric magnetic field dependence 
of the thermopower in Fig.~\ref{fig:exp}a. In this paper we develop a
general analytical approach that explains both antisymmetric and symmetric 
dependence of the thermopower observed in the experiment. 
Our theory is based on the quasiclassical kinetic equations 
in the limit of weak proximity effect. Even though this approximation 
assumes a small transparency of the normal-metal/superconductor 
interfaces in Fig.~\ref{fig:element}, it does not affect the symmetry
of the obtained results. 

In the Section~\ref{sec:para} we solve the kinetic equations 
for the parallelogram interferometer
depicted schematically in Fig.~\ref{fig:para}.
In accordance with earlier studies we find that 
the charging of the superconducting arm of the interferometer 
is determined by the interplay of the supercurrent and the temperature gradient. 
The formation of the charge imbalance is analogous 
to that in the Pethick-Smith effect.\cite{Pethick1979} 
The extra charge accumulated by the superconductor is, therefore, 
a purely antisymmetric function of both the temperature gradient 
and the applied magnetic flux. In this case the left-right symmetry 
of the structure has to be broken in order to observe the effect 
of proximity-induced correlations on the thermopower of the 
normal-metal wire.\cite{Virtanen2004a}  
The symmetry breaking can be caused either by 
a difference between $L_1$ and $L_2$
in Fig.~\ref{fig:para} assuming that these 
distances are much smaller than the phase-coherence 
length $L_\phi$, or by a difference
in the $NS$ interface transparencies $\alpha_1$ and $\alpha_2$, 
which are given by the ratio of the normal wire resistance 
per unit length to the interface resistance. In the latter case 
the phase coherence of quasiparticles 
in the normal-metal wire is irrelevant for the symmetry breaking. 
We also stress that the energy relaxation processes 
in the superconductor play an important role in the theory of
the thermoelectric effect.
Indeed, the time-independent solution to the kinetic problem 
exists only if the detailed balance condition at a given energy is broken. 
Another words, in the steady-state limit, only the total charge 
transmitted through the superconducting wire 
is conserved while the energy density of the charge flow is not. 
The steady-state regime develops at times exceeding the 
imbalance relaxation time $\tau_Q$ inside the superconducting wire.
The resulting behavior of the thermopower in the parallelogram interferometer
is illustrated in Fig.~\ref{fig:Yplot} as a function of temperature.

In the Section~\ref{sec:house} we apply the same approach to
the house interferometer, which is shown 
schematically in Fig.~\ref{fig:house}.  
In sharp contrast to the previous case we find a finite 
contribution to the thermopower in the absence of a charge imbalance 
between the normal-metal and the superconducting branches. 
Such contribution has an antisymmetric dependence on the magnetic flux 
and requires a left-right asymmetry of the wire $N'$ 
that must originate in a difference between $d_1$ and $d_2$ rather than 
in a difference between the $NS$ interface transparencies $\tilde{\alpha}_1$ and
$\tilde{\alpha}_2$. This asymmetric dependence 
of the thermopower in the house interferometer is a phase-coherent phenomena
that can be attributed to the interference of quasiparticle trajectories 
shown in Fig.~\ref{fig:track}. Apart from the evident requirement $d_1,d_2\ll L_\phi$ 
this effect is strongly sensitive to the position of the Andreev reflection.\cite{Stoof1996} 
In the parallelogram interferometer the interference contribution to the thermopower
is averaged to zero since it involves the summation over the trajectories of different lengths.  
 
The odd oscillations of the thermopower are, however, not seen 
in experiments with the house interferometer. Instead, 
a sign-definite thermopower, which has an even dependence on the
magnetic flux $\Phi$, has been observed.\cite{Eom1998,Parsons2003,Cadden2007} 
Moreover, the maximal values of the symmetric thermopower in Fig.~\ref{fig:exp}b
correspond to $\Phi=n\Phi_0$, where $n$ is an integer number 
and $\Phi_0=h/2e$ is the flux quantum. We demonstrate that 
this behavior is a signature of the finite charge imbalance 
maintained between the Cooper-pair chemical potential in the superconductor 
and the quasiparticle chemical potential in the normal-metal. 
Our theory predicts that the thermopower in this case is proportional to  
$\mu\lt(1+\cos2\pi\Phi/\Phi_0\rt)$, hence its magnetic field dependence 
agrees with the experimental data\cite{Eom1998} in Fig~\ref{fig:exp}b.
The thermopower determined by this effect has a peculiar sign-reversal 
behavior with increasing temperature that is illustrated in Fig.~\ref{fig:Hs}.
This temperature dependence is in qualitative agreement with the experiments 
by Parsons {\it et al.}\cite{Parsons2003}

To our opinion the experiments of Refs.~\onlinecite{Eom1998,Parsons2003,Cadden2007} 
provide a clear signature of a finite steady-state imbalance $\mu$ 
in the house interferometer. We demonstrate that a tiny chemical potential 
imbalance between different branches of the house interferometer
has a great impact on the thermopower due to the proximity effect.
The possible origin of the tiny imbalance is in the 
usual thermoelectric effect caused by a heat dissipation 
in the second normal-metal wire $N'$. The validity of this scenario can be tested 
by varying the chemical potential in $N'$ with the help of an additional electrode 
or by a detailed measurement of the temperature dependence of the thermopower
in the house interferometer.
 
The remainder of the paper is organized as follows. 
In the Section~\ref{sec:kin} we introduce the well-known 
quasiclassical approach to the proximity effects 
in diffusive mesoscopic wires. In the Section~\ref{sec:boundary} 
the rigid boundary conditions at a three terminal contact 
are given. The Sections~\ref{sec:para}, \ref{sec:house} 
are devoted to the analytical calculation of the thermopower 
for the parallelogram and the house interferometer, correspondingly. 
We summarize our conclusions in the Section~\ref{sec:conclusion}.  

\section{quasiclassical theory}
\label{sec:kin}

Andreev reflection is the main microscopic mechanism of charge transport 
between the superconductor (S) and the normal-metal (N) at low temperatures.
It can be seen from the normal-metal side as the conversion of electron-like
excitations of the energy $\ep$ to the hole-like ones. The reflection takes place 
at the $NS$ boundary and is responsible for the phase coherence of electrons 
and holes at a distance $\sqrt{D/\ep}$ in the diffusive normal metal, 
where $D$ is the diffusion coefficient (we set $\hbar=k_B=1$ throughout the rest of the paper). 
From the superconductor side the same process can be viewed as a diffusion of Cooper pairs
that brings the superconducting correlations into the normal metal.

The characteristic scale of the proximity effect at a temperature $T$ 
is determined by the coherence length $\xi=\sqrt{D/T}$. 
The electron-hole coherence 
can be detected in the diffusive metal 
provided $T\lesssim E_c$, where $E_c=D/L^2$ 
is the Thouless energy associated with the distance $L$ to the $NS$ interface. 
On the ballistic scales, $E_c > \Delta$, the role of $E_c$ is played by $\Delta$,
the absolute value  of the superconductor energy gap.  

The conditions for the diffusive proximity effect were fulfilled in the pioneering experiments 
with Andreev interferometers performed by Chandrasekhar {\it et al.}\cite{Eom1998} and 
Parsons {\it et al.}\cite{Parsons2003} 
We, therefore, reduce our consideration to the diffusive approximation 
of the quasiclassical theory that is described by the Usadel equation.

We shall start from the stationary Usadel equation written for 
the quasiclassical Green's function $\check{g}(x,\ep)$ 
in a diffusive normal-metal wire, 
\be
\label{Usadel}
D\frac{d \check{I}}{dx}+[i \ep \sigma_z, \check{g}]=0,
\quad \check{I}\equiv \check{g}\frac{d \check{g}}{dx},
\e 
where $D$ is the diffusion coefficient, $\check{I}$ is the matrix current, and
$x\in(-L_1,L_2)$ is the coordinate along the normal-metal wire. 
The Green's function is represented by a 
matrix in the Keldysh space, 
\be
\label{quasi}
\check{g}=\lt(\ba{cc}\hat{g}^R& \hat{g}^K\\ 0& \hat{g}^A \ea\rt), 
\qquad \check{g}^2=1,
\e
which yields the quasiclassical constraint $\check{g}^2=1$. 

Despite the absence 
of the superconducting order parameter in the metallic wire, the proximity effect 
will lead to a superconducting correlations that are described by the anomalous
components $f^{R,A}$ of the Green's function. These correlations are taken into account 
by the extension to the electron-hole space.  The spectral (retarded and advanced) sector 
of the Green's function in this space is parameterized as
\be
\label{prmRA}
\hat{g}^{R (A)} =\lt(\ba{cc} \pm g^{R (A)} & f^{R (A)} \\ 
\bar{f}^{R (A)} & \mp g^{R (A)} \ea\rt), 
\e
where bar stands for the charge conjugation. The relation
\be
\qquad g^{R (A)}=\sqrt{1-f^{R (A)} \bar{f}^{R (A)}},
\e
follows from the quasiclassical constraint (\ref{quasi}). 
The Keldysh component of the constraint, 
$\hat{g}^R \hat{g}^K+ \hat{g}^K \hat{g}^A =0$, 
suggests that $\hat{g}^K$ has only two linearly independent entities. 
In what follows we use the standard parameterization 
\be
\label{prmK}
\hat{g}^K=\hat{g}^R \hat{h}-\hat{h}\hat{g}^A, 
\e
where the diagonal matrix 
\be
\hat{h}=h+\sigma_z h_\sigma, 
\e 
has two degrees of freedom: $h$  and $h_\sigma$. 

We model the experimental situation by an idealized system consisting of 
two equilibrium reservoirs connected by a non-interacting normal-metal diffusive wire $N$.  
The distribution functions $\hat{h}_{1}$ and $\hat{h}_2$ in the reservoirs
are parameterized by the chemical potentials $\mu_1$, $\mu_2$ 
and the temperatures $T_1$, $T_2$ as
\be
\label{h_conditions}
\lt.\hat{h}\rt|_{x\to -L_1,L_2}=
\lt(\ba{cc} h_{1,2} & 0\\ 0 & \bar{h}_{1,2} \ea \rt), 
\e
where
\be
h_a=\tanh\frac{\ep-\mu_a}{2T_a}, \qquad
\bar{h}_a=\tanh\frac{\ep+\mu_a}{2T_a},
\e
and $a=1,2$. The functions $(1-h_a)/2$, $(1-\bar{h}_a)/2$ 
are the Fermi distribution functions of electrons and holes, 
correspondingly. The electric and the heat current in the 
normal-metal wire are obtained in the Usadel approximation 
from the Keldysh component of the matrix current $\check{I}$, 
\beml
\label{currents}
\beq
&&J^e=\frac{e D \nu}{4} \int\!\! d\ep\, j^e(\ep), \qquad  j^e(\ep)=\tr \sigma_z \hat{I}^K,
\label{Je}\\
&&J^Q=\frac{D \nu}{4}\int\!\! d\ep\, \ep\,j^Q(\ep), \qquad j^Q(\ep)=\tr \hat{I}^K,
\label{JQ}
\eq
\eml
where $\hat{I}^K=\hat{g}^R (d\hat{g}^K/dx) + \hat{g}^K (d\hat{g}^A/dx)$.

The Usadel equation (\ref{Usadel}) does not take into account inelastic processes,
hence the charge and heat flow is conserved for each energy.
Indeed, from the diagonal components of its Keldysh sector we obtain 
the conservation laws
\be
\label{kinetic}
\frac{d j^e(\ep)}{dx}=\frac{dj^Q(\ep)}{dx}=0,
\e
which play a role of the kinetic equation in the normal-metal wire.
The diagonal components of the retarded and advanced component of the 
Usadel equation (\ref{Usadel}) imply the conservation of the "spectral" current
\be
\label{Wronskian}
\frac{d W^R}{dx}=\frac{d W^A}{dx}=0,
\e
where $W^{R}=f^{R}(d\bar{f}^{R}/dx)-\bar{f}^{R}(df^{R}/dx)$. This 
condition is equivalent to the conservation of the supercurrent, which can 
flow in the normal metal in the presence of the proximity effect.

From Eqs.~(\ref{currents}) we find the current energy densities $j^e(\ep)$, $j^Q(\ep)$ 
in the parameterization of Eqs.~(\ref{prmRA},\ref{prmK}) as
\beml
\label{current_density}
\beq
\label{current_e}
j^e(\ep)&=&M_+\frac{dh_\sigma}{dx}-U\frac{dh}{dx}+Wh,\\
\label{current_Q}
j^Q(\ep)&=&M_-\frac{dh}{dx}+U\frac{dh_\sigma}{dx}+Wh_\sigma,
\eq
\eml 
where the following functions, which depend on the energy and coordinate, 
are introduced
\beml
\beq
&&M_{\pm}=\Big(2(1+g^Rg^A)\pm (f^R\bar{f}^A+f^A\bar{f}^R)\Big),\\
&&U=f^R\bar{f}^A-f^A\bar{f}^R,\quad W=W^R-W^A.
\eq
\eml
Thus, the kinetic part of the transport problem 
is fully described by Eqs.~(\ref{kinetic},\ref{current_density}). 

It is instructive to apply this formalism to the transport in a metallic wire
in the absence of a proximity effect. In this case one finds $f^{R (A)}=0$, hence 
$g^R=-g^A=\sigma_z$ and
\be
\label{noproximity}
j^e=4\frac{dh_\sigma}{dx},\qquad j^Q=4\frac{dh}{dx}.
\e
The solution to Eqs.~(\ref{kinetic}), which yields the boundary conditions
(\ref{h_conditions}), reads
\beml
\label{linear}
\beq
h(\ep,x)&=&\frac{1}{4}\lt(h_0+\frac{L_2-L_1-2 x}{L} h_T\rt), \\
h_\sigma(\ep,x)&=&\frac{1}{4}\lt(h_\mu+\frac{L_2-L_1-2 x}{L} h_{\mu T}\rt), 
\eq
\eml
where the following notations are used 
\beml
\label{h_functions}
\beq
h_0&=&h_1+\bar{h}_1+h_2+\bar{h}_2,\\
h_T&=&h_1+\bar{h}_1-h_2-\bar{h}_2,\\ 
h_\mu&=&h_1-\bar{h}_1+h_2-\bar{h}_2,\\
h_{\mu T}&=&h_1-\bar{h}_1-h_2+\bar{h}_2.
\eq
\eml
Substituting  Eqs.~(\ref{linear}) to Eqs.~(\ref{noproximity}) we obtain 
$j^e=-(2/L)h_{\mu T}$ and $j^Q=-(2/L)h_{T}$. 

If a temperature gradient is applied between the reservoirs 
that are kept at the same chemical potential $\mu=\mu_1=\mu_2$, 
the integration in Eq.~(\ref{Je}) gives $J^e=0$, hence $\eta=0$.
The heat current in the limit $|T_1-T_2|\ll T$ is given by Eq.~(\ref{JQ})
as
\be
J^Q=-\kappa (T_2 - T_1), \qquad \kappa=\frac{\pi^2}{3} \frac{2 D \nu T}{L},
\e 
where $T=(T_1+T_2)/2$ is the mean temperature in the system and $\kappa$
is the thermal conductivity.

Similarly, if the chemical potential difference is applied between 
the reservoirs that are kept at the same temperature $T$,
the integration in Eqs.~(\ref{currents}) gives $j^Q=0$ and
\be
\label{sigma}
J^e=\sigma \frac{\mu_2-\mu_1}{(-e) L}, \quad \sigma= 2 e^2 D \nu,
\e
where $\sigma$ is the Drude result for the electric conductivity.
 
We shall stress that the vanishing of the thermoelectric coefficients $\eta$ and $\zeta$
in the calculation above is a consequence of the linearization of the quasiparticle 
spectrum near the Fermi energy $\ep_F$.  Such a linearization is essential 
for the quasiclassical approximation and leads to the exact electron-hole symmetry in metals. 
The proximity effect can, however, break this symmetry through the possible
energy dependence of $M_{\pm}$, $U$ and $W$ in Eq. ~(\ref{current_density})
that can give rise to a finite thermoelectric response in the system.
Such thermoelectric phenomena can be analyzed 
within the quasiclassical approximation.

Since the superconducting arm of the interferometer used in experiments 
exceeds the electron-phonon scattering length it can be described 
by the equilibrium quasiclassical Green's function $\check{g}_S$.
We assume for simplicity that the superconducting energy gap drops abruptly to zero 
in the normal-metal/superconductor interface. Even though this assumption 
is not self-consistent and disregards a small suppression of $\Delta$ near the $NS$ boundary, 
it does not affect our conclusions on the symmetry of the thermopower oscillations.

We, therefore, describe the superconductor
by the standard bulk expressions for the spectral components of $\check{g}_S$
that are given by 
\be
\label{gS}
\hat{g}^{R(A)}_S=\frac{1}{\sqrt{(0\mp i \ep)^2+\Delta^2}}
\lt(\ba{cc} -i\ep & \Delta e^{i\chi} \\ \Delta e^{-i \chi} & i\ep\ea\rt),
\e
where $\Delta \exp(i\chi)$ is the superconducting order parameter.

The external magnetic flux $\Phi$ piercing the loop leads to a gradient 
of the phase $\chi$, hence the phase acquires
different values $\chi_1$ and $\chi_2$ at the opposite ends 
of the superconducting wire. The order parameter phase difference, 
\be
\varphi=\chi_1-\chi_2,
\e 
is related to the magnetic flux as $\varphi=2 \pi \Phi/\Phi_0$.
The Keldysh component of $\check{g}_S$ is parameterized by
\be
\hat{g}^K_S=(\hat{g}^R_S -\hat{g}^A_S)h_S,  
\e
where $\hat{h}_S=\tanh(\ep/2T_S)$ is the quasiparticle distribution function
and $T_S$ is the temperature of the superconductor. Here we choose 
the chemical potential of the Cooper pairs as zero and set $h_{S\sigma}=0$. 
For $\ep<\Delta$ one finds from Eq.~(\ref{gS}) that $\hat{g}^R_S=\hat{g}^A_S$, 
hence $\hat{g}^K_S=0$.  In this paper we are concerned with ultra
low temperatures $T\ll \Delta$ and , therefore, disregard all quasiparticle effects 
in the superconductor. 

The effective energy scale for the proximity effect in the normal metal is given 
by the Thouless energy $E_c=D/L^2$, where $L$ is the distance 
from the $NS$ interface.  To describe the experimentally relevant situation
$T\sim E_c \ll \Delta$, we restrict our consideration to the excitation energies 
that are much smaller than the absolute value of the superconducting gap, therefore
\be
\hat{g}_S^R=\hat{g}_S^A=\lt(\ba{cc} 0 & e^{i\chi}\\ e^{-i\chi}& 0\ea\rt).
\e
We will see below that our results for the thermopower are indeed defined 
by small excitation energies $\ep\leq E_c$, which proves the   
the consistency of this approximation. 

In addition we assume that the proximity effect is weak due to a low transparency 
of the NS interfaces. This assumption allows for the linearization 
of the Usadel equation (\ref{Usadel}) with respect to the anomalous component
of the Green's function,
\be
\label{UsadelfR}
\frac{d^2f^{R}}{dx^2}=z^2 f^{R}, \qquad z^2\equiv\frac{2(0 - i\ep)}{D},
\e
where we replace $\ep$ by $\ep - i0$ in order to stress the 
analytical properties of $f^R$. The solution for the advanced Green's function
is obtained by the substitution $\ep\to-\ep$. The conservation of the supercurrent 
in Eq.~(\ref{Wronskian}) is equivalent in this approximation 
to the standard property of the Wronskian of a linear second-order 
differential equation. The following symmetry relations simplify
the subsequent analysis,
\be
\label{symmetries}
f^A(\ep)=f^R(-\ep),\> \bar{f}^R(\Phi)=f^R(-\Phi),\> \bar{f}^A=(f^R)^*.
\e

\section{boundary conditions}
\label{sec:boundary}

\begin{figure}[t]
\includegraphics[width=0.7\columnwidth]{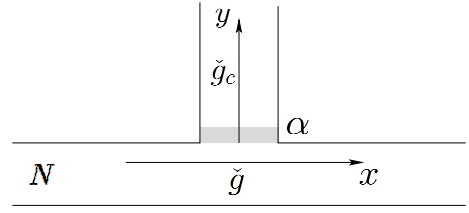}
\caption{
The three-terminal junction formed by a contact attached to the main 
normal-metal wire $N$. The generalized rigid boundary conditions 
at the junction are given by Eqs.~(\ref{cond},\ref{der_cond}) in the matrix notation. 
The Green's function $\check{g}$ is continuous along the wire $N$ 
but there is a jump in the Green's function across the tunnel barrier
so that $\check g\neq \check g$ at the junction. 
The kinetic component of Eq.~(\ref{der_cond}) is 
equivalent to Eqs.~(\ref{je_junction},\ref{jQ_junction}). 
}
\label{fig:element}
\end{figure}

The boundary condition for a junction shown in Fig.~\ref{fig:element}
is an important ingredient of our description of the proximity effect 
in Andreev interferometers. We assume that a contact wire, which is 
either a superconductor in the parallelogram interferometer 
or a normal-metal wire $N'$ in the house interferometer, is described 
by the Green's function $\check{g}_c$. The Green's function 
in the main wire $N$ is denoted by $\check{g}$. 

The transparency of the barrier between the contact and 
the normal-metal wire is parameterized by a coefficient $\alpha=T_B/\ell_B$.
It has a dimension of an inverse length, where $T_B$
is the barrier transmission probability per channel 
and $\ell_B$ is an effective barrier length, which is 
of the order of the mean free path in $N$. 
The coefficient $\alpha$ can also be regarded as the ratio of 
the normal-wire resistance per unit length to the interface resistance
and has a dimension of an inverse length.

The generalized "rigid" boundary conditions\cite{GKIrev} apply 
provided $T_B\ll 1$ for each transmission channel at the junction.
We stress that the tunnel barrier in Fig.~\ref{fig:element}
is placed between the normal-metal wire $N$ and the contact wire.
As the result the green's function $\check{g}$ is continuous along 
the normal metal wire while its derivative has a discontinuity that 
is determined by a difference between $\check{g}$ and $\check{g}_c$.
The continuity of the Green's function in the $N$ wire is
formally written as
\be
\label{cond}
\delta\lt[\check{g}\rt]_{x_0}\equiv
\lim_{\delta\to 0}
\Big(\check{g}(x_0+\delta)-\check{g}(x_0-\delta)\Big)=0.
\e
The discontinuity of the derivative of $\check{g}(x)$ is obtained from the
matrix current conservation in the three terminal junction
\be
\label{der_cond}
\delta\lt[\check{I}\rt]_{x_0}=
\lt.\frac{\alpha}{2}[\check{g}_c,\check{g}]\rt|_{x_0}.
\e
Note, that the green's function is discontinuous across 
the tunnel barrier, i.e. $\check{g} \neq \check{g}_c$ at $x=x_0$.
One can easily demonstrate that Eq.~(\ref{der_cond}) is a discrete
analog of the Usadel equation (\ref{Usadel}) and
yields the same set of conservation laws. 

For instance, the spectral part of Eqs.~(\ref{cond},\ref{der_cond}) leads to
\beml
\label{fR_boundary}
\beq
&&\delta\lt[W\rt]_{x_0}=\tilde{W}=\tilde{W}^R-\tilde{W}^A,\\
&&\tilde{W}^R=-\alpha\lt( f^R\bar{f}^R_c-f^R_c\bar{f}^R\rt),
\eq
\eml
which illustrates the conservation of the suppercurrent at the junction.

In a full analogy with Eqs.~(\ref{kinetic},\ref{current_density}) 
the Keldysh sector of Eqs.~(\ref{cond},\ref{der_cond}) is given by
\beq
&&\!\!\!\!\delta[j^e(\ep)]_{x_0}=\frac{\alpha}{2}\tr \sigma_z[\check{g}_c,\check{g}]^K_{x_0} \n\\
&&=\alpha (h_\sigma\!-\!h_{c\sigma}) \tilde{M}_+ - \alpha (h\!-\!h_{c})\tilde{U}
+\frac{h\!+\!h_c}{2} \tilde{W},
\label{je_junction}
\\
&&\!\!\!\!\delta[j^Q(\ep)]_{x_0}=\frac{\alpha}{2}\tr [\check{g}_c,\check{g}]^K_{x_0}\n\\
&&=\alpha (h\!-\!h_{c}) \tilde{M}_- +\alpha (h_\sigma\!-\!h_{c\sigma})\tilde{U}
+\frac{h_\sigma+h_{\sigma c}}{2} \tilde{W},\mbox{\hspace*{0.2cm}}
\label{jQ_junction}
\eq
where the following notations are introduced
\beq
\tilde{M}_{\pm}&=&(g^R\!+\!g^A)(g_c^R\!+\!g_c^A)+\frac{1}{2}(f^R\!+\!f^A)(\bar{f}^R_c\!+\!\bar{f}^A_c)\n\\
&&+\frac{1}{2}(\bar{f}^R\!+\!\bar{f}^A)(f^R_c\!+\!f^A_c),\\
\tilde{U}&=&\frac{1}{2}(f^R \bar{f}^A_c\!-\!\bar{f}^Rf^A_c-f^A\bar{f}^R_c+\bar{f}^Af^R_c),\\
\tilde{W}&=&-\alpha(f^R \bar{f}^R_c\!-\!\bar{f}^Rf^R_c-f^A\bar{f}^A_c+\bar{f}^Af^A_c).
\eq
The Equations (\ref{je_junction},\ref{jQ_junction}) relate the discontinuity 
in the current in the normal-metal wire $N$ to the current flowing into the contact wire.

 
\begin{figure}[t]
\includegraphics[width=0.9\columnwidth]{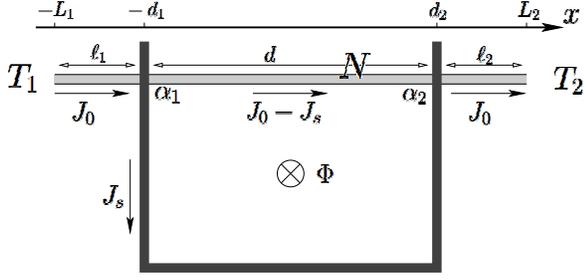}
\caption{
The parallelogram interferometer. Diffusive normal-metal 
wire $N$ of the length $L=L_1+L_2$ connects the temperature reservoirs 
denoted as $T_1$ and $T_2$.  The superconducting wire (dark line)
is contacting $N$ at two junctions separated by a distance $d$.
The $NS$ interfaces are characterized by the 
transparency parameters $\alpha_1$ and $\alpha_2$. Only antisymmetric
dependence of the thermopower on the magnetic flux is allowed by symmetry.
The effect requires $\ell_1\neq \ell_2$ or $\alpha_1\neq \alpha_2$.   
}
\label{fig:para}
\end{figure}

\section{thermopower in the parallelogram interferometer}
\label{sec:para}

The parallelogram interferometer realized in the experiment of Ref.~\onlinecite{Eom1998} 
is depicted schematically in Fig.~\ref{fig:para}. The thermopower is measured 
between the temperature reservoirs $T_1$ and $T_2$ that are 
connected by the normal-metal wire $N$ (Au) of the length $L=L_1+L_2\sim 1.7 \mu m$.
The wire is contacted to a long superconducting arm at the points $x=-d_1$ and 
$x=d_2$ with a distance $d=d_1+d_2 \sim 0.2 \mu m$ between them.
The thermopower measurement in Fig.~\ref{fig:exp}a is performed at 
the temperature $T=350mK$. The superconducting coherence length
at this temperature is estimated as $\xi=\sqrt{D/T}\sim 0.54  \mu m$, while 
the phase coherence length is approximately given by $L_\phi\sim 3.5 \mu m $.
Thus, the experimental parameters are such that $d\lesssim \xi$. Therefore, 
if an external magnetic flux $\Phi$ is piercing the interferometer,
the supercurrent $J_s$ flows in the normal-metal wire between the $NS$ junctions. 
 
In the subsequent analysis we let $\mu_1=\mu_2=\mu$ 
and calculate the quasiparticle current $J_0$ between the reservoirs 
to the first order in the temperature difference $T_1-T_2$, 
thus, obtaining the coefficient $\eta$. The thermopower is, then, calculated
from the relation $S=\eta/\sigma$. In the limit of weak proximity effect 
the Drude result for $\sigma$ can be used. We note that a similar approach has been 
developed in the series of publications by Virtanen 
and Heikilla.\cite{Virtanen2004a,Virtanen2004b,Virtanen2007}  

In order to keep our formulas compact we define the distances $\ell_1$
and $\ell_2$ between the $NS$ interfaces and  the corresponding reservoirs 
as $\ell_a\equiv L_a-d_a$, where $a=1,2$. The important simplification,  
which enables us to solve the problem analytically, 
is the linearization of the Usadel equations (\ref{UsadelfR}) for the spectral 
components of the Green's function $\check{g}(x)$. 
This approximation is justified since the wire is in a good contact 
to the reservoirs with a vanishing proximity effect, 
\be
\label{fra_zero}
\lt.f^{R(A)}\rt|_{x=-L_1}=\lt.f^{R(A)}\rt|_{x=L_2}=0.
\e
We remind that the chemical potential $\mu$ in the reservoirs is measured 
with respect to that of the Cooper pairs in the superconductor, 
therefore, the temperature equilibrium is characterized by $\mu=0$.  

We start by pointing out that the energy relaxation processes are 
important for the existence of a steady-state regime with a finite $\mu$
in the parallelogram interferometer. Indeed, if scattering 
is purely elastic the charge conservation law in the supercondutor 
must hold for each energy. In this case the time-independent
solution to the kinetic equation does not exist. Following 
other works\cite{Seviour2000,Virtanen2004a} we relax 
the energy-resolved condition to the conservation 
of a total charge in the superconductor that is given by 
\be
\label{imb_cond}
\int d\ep\, \lt( \delta\Big[j^e\Big]_{d_2}+\delta\Big[j^e\Big]_{-d_1}\rt)=0,
\e
in the notations of Eq.~(\ref{cond}). The charge conservation condition in this form
can be regarded as the equation on the imbalance $\mu$. 
A finite supercurrent $J_s$ in the loop leads to a non-zero imbalance, which 
may act as a source of an electron-hole symmetry violation in the system.
In the analogy with the Pethick-Smith effect,\cite{Pethick1979} 
the imbalance defined by Eq.~(\ref{imb_cond}) is proportional 
to the scalar product of the temperature gradient and the supercurrent. 
We will see below that  $\mu= A\,(T_1-T_2)\sin 2\pi \Phi/\Phi_0$, 
where $A$ is a dimensionless coefficient that is mainly 
determined by the ratio of the Thouless energy $E_c=D/L^2$ to 
the mean temperature $T=(T_1+T_2)/2$, such that  $A$ is maximal
for $T\sim E_c$.

From Eq.~(\ref{imb_cond}) we express 
the current in the superconducting wire as
\be
\label{Js}
J_s=\frac{\sigma}{16 e} \int d\ep\, 
\lt(\delta\Big[j^e\Big]_{d_2}- \delta\Big[j^e \Big]_{-d_1}\rt),
\e
where $\sigma$ is the Drude conductivity given by Eq.~(\ref{sigma}).
Furthermore, the spacial integration of the electric current along 
the normal-metal wire gives 
\be
\label{J0}
J_0=\frac{d}{L}J_s^e+\frac{1}{L}\int_{-L_1}^{L_2}dx\,\,\frac{\sigma}{8e}\int d\ep\, j^e.
\e
Thus, the boundary condition (\ref{je_junction}) at the $NS$ interfaces together 
with Eqs.~(\ref{currents},\ref{imb_cond},\ref{Js},\ref{J0}) describe
the kinetic part of the problem.

Let us now perform the calculation of $J_0$ to the second order 
in the parameters $\alpha_{1,2}$. In this approximation the anomalous components 
of the Green's function, $f^{R(A)}$, are found from the linearized 
Usadel equation (\ref{UsadelfR}) with the help of the boundary conditions 
(\ref{fra_zero}) and 
\be
\label{gradf}
\delta\lt[\frac{df^R}{dx}\rt]_{-d_1} = -\alpha_1 e^{i \chi_1},\quad
\delta\lt[\frac{df^R}{dx}\rt]_{d_2} = -\alpha_2 e^{i \chi_2},
\e
where the coefficients $\alpha_{1}$ and $\alpha_2$ introduced in the 
previous Section are determined by the $NS$ interface transparency.   
The conditions (\ref{gradf}) follow directly from 
the retarded component of Eq.~(\ref{der_cond}). 
With the help of the following notations
\beml
\beq
F_{a} &=&  \alpha_a^2 \frac{\sinh z\ell_a\, \sinh z(L-\ell_a)}{z L \sinh z L},\\
F_{12}&=&\alpha_1\alpha_2 \frac{\sinh z\ell_1\, \sinh z\ell_2}{z L \sinh z L},
\eq
\eml
we express the anomalous Green's function near the $NS$ interfaces as
\beml
\label{fRfA}
\beq
\frac{\alpha_1}{L} f^R(-d_1) &=& e^{i \chi_1} F_1+ e^{i \chi_2} F_{12},
\\ 
\frac{\alpha_2}{L} f^R(d_2) &=& e^{i \chi_2} F_2+ e^{i \chi_1} F_{12}, 
\eq
\eml
and find the supercurrent density in the normal-metal wire as
\be
W=\lt\{\ba{ccl}
W_0 &\quad & x\in (-d_1,d_2),\\
0 &\quad & x\notin (-d_1,d_2) ,
\ea \rt.
\e
where $W_0=-4 L \sin \varphi\, \im F_{12}$ . 
From Eq.~(\ref{current_e}) we obtain
\beml
\beq
\delta[j^e]_{-d_1} &=& h(-d_1)\, W_0+4
\delta\lt[\frac{dh_\sigma}{dx}\rt]_{-d_1},\\
\delta[j^e]_{d_2} &=& -h(d_2)\, W_0+4
\delta\lt[\frac{dh_\sigma}{dx}\rt]_{d_2},
\eq
\eml
to the second order in $\alpha_{1,2}$. 

The charge conservation condition (\ref{imb_cond}), 
which determines the voltage difference $\mu/e$ 
between the superconductor and the normal-metal wire, can be written as
\be
\label{s1}
\frac{d}{L}\int\!\! d\ep\, h_T W_0=
-8\!\! \int\!\! d\ep \Big\{ \delta\lt[\frac{dh_\sigma}{dx}\rt]_{d_2}
\!\!\!+\delta\lt[\frac{dh_\sigma}{dx}\rt]_{-d_1}\Big\}.
\e
The left-hand side of this expression is proportional to the
temperature gradient due to the distribution function $h_T$.
It is also antisymmetric with respect to the order parameter phase difference
$\varphi=\chi_1-\chi_2=2\pi \Phi/\Phi_0$ due to $W_0$. The right-hand side 
of Eq.~(\ref{s1}) is proportional to the imbalance $\mu$. 
We, therefore, come to the conclusion that the steady-state imbalance
is an odd function of the temperature gradient and the applied flux.

From Eq.~(\ref{Js}) we readily find the supercurrent
\beq
J_s &=&\frac{\sigma}{8e}\int d\ep
\lt\{-\frac{1}{4}\lt(h_0+\frac{\ell_2-\ell_1}{L}h_T\rt)W_0\rt. \n\\
&+&\lt.2\lt( \delta\lt[\frac{dh_\sigma}{dx}\rt]_{d_2}-
\delta\lt[\frac{dh_\sigma}{dx}\rt]_{-d_1} \rt)
\rt\}.
\label{s2}
\eq
We also simplify Eq.~(\ref{J0}) by calculating 
the spatial integral from the first term in the right-hand side
of Eq.~(\ref{current_e}) to the leading order in $\alpha$
\be
\frac{1}{L}\int_{-L_1}^{L_2} h W dx= \frac{d}{4L}\lt(h_0+\frac{\ell_2-\ell_1}{L}h_T\rt)W_0.
\e
Thus, we obtain
\beq
J_0&=&\frac{\sigma}{8e}\int d\ep
\Bigg\{\frac{2d}{L}
\lt( \delta\lt[\frac{dh_\sigma}{dx}\rt]_{d_2}-
\delta\lt[\frac{dh_\sigma}{dx}\rt]_{-d_1} \rt)\n\\
&+&\frac{h_{T}}{2L}\frac{1}{L}\il_{-L_1}^{L_2}
dx\,U -\frac{h_{\mu T}}{2L}\frac{1}{L}\il_{-L_1}^{L_2}
dx\,M_+\Bigg\},
\label{s3}
\eq
where the spatial derivatives of $h$ and $h_\sigma$ 
are found from Eqs.~(\ref{linear}). 

The functions $h_{\mu}$ and $h_{\mu T}$ introduced in 
Eq.~(\ref{h_functions}) are symmetric with respect to the energy $\ep$ and 
antisymmetric with respect to the imbalance $\mu$, while $h_{0}$ and $h_{T}$
are antisymmetric in $\ep$ and symmetric in $\mu$. 
Moreover, the last term in Eq.~(\ref{s3}) can be disregarded in the 
linear response analysis. Indeed, the distribution function $h_{\mu T}$
is of a second order in the temperature gradient due to the fact that 
$\mu$ found from Eq.~(\ref{s1}) is itself proportional to the temperature 
gradient. We shall, therefore, omit all terms containing $h_{\mu T}$ in 
our analysis of the thermopower for the parallelogram interferometer. 
We note, however, that the last term in Eq.~(\ref{s3}) is 
the only one that changes sign with increasing temperature. 
We, therefore, regard the sign-reversal behavior of the thermoelectric response 
of the parallelogram interferometer observed by Parsons {\it et al.}\cite{Parsons2003} 
as the second order effect that is beyond the linear response.
Unlike other terms in Eq.~(\ref{s3}) the last one is finite for a left-right symmetric setup
and can easily dominate an experimental measurement for 
any finite temperature gradient. We postpone the detailed discussion of this 
contribution to the next Section. 

Omitting the terms containing $h_{\mu T}$ we find that
the discontinuity of the derivative $dh_\sigma/dx$ near the $NS$ 
interfaces is determined by the boundary conditions (\ref{je_junction}) as
\beml
\label{der_a}
\beq
\delta\lt[\frac{dh_\sigma}{dx}\rt]_{-d_1}&=&
\frac{L}{4}\, h_\mu \re ( F_1+F_{12}\cos\varphi ), \\
\delta\lt[\frac{dh_\sigma}{dx}\rt]_{d_2}&=&
\frac{L}{4}\, h_\mu \re ( F_2+F_{12}\cos\varphi),
\eq
\eml
where we take advantage of the condition $h_{S\sigma}=0$.

The spatial integrals in Eq.~(\ref{s3}) can be expressed through 
the anomalous Green's functions $f^{R,A}$ at the $NS$ interfaces with 
the help of Eq.~(\ref{UsadelfR}) and the integration by parts. 
With the help of Eqs.~(\ref{gradf},\ref{fRfA}) we obtain
\be
\label{int}
\frac{\ep}{D}\il_{-L_1}^{L_2}dx\, f^R \bar{f}^A=
L \im \lt(\frac{F_1+F_2}{2}+F_{12}\cos\varphi\rt),
\e
that leads to 
\be
\label{zero}
\frac{1}{L}\int_{-L_1}^{L_2}dx\,(f^R\bar{f}^A-f^A\bar{f}^R)=0.
\e
Thus, the second term in Eq.~(\ref{s3}) is also vanishing.
We will see in the next Section that this term is responsible for the 
particle-hole interference contribution to the thermopower that
can be finite in the house interferometer. Finally, the first term in Eq.~(\ref{s3}) 
gives rise to the only non-vanishing contribution to the current $J_0$ 
that is linear with respect to the temperature gradient. 

From Eqs.~(\ref{der_a}) we obtain
\be
\label{J0final}
J_0=-\frac{\sigma d}{8 e}\int d\ep\, h_{\mu} \re \frac{F_1-F_2}{2},
\e
therefore the quasiclassical thermoelectric effect is absent in the left-right 
symmetric device $F_1=F_2$ in accordance with Ref.~\onlinecite{Seviour2000}
and general symmetry considerations. The thermopower is 
readily found from the relation $S=(J_0/\sigma)(T_2-T_1)^{-1}$, where 
we disregard the corrections to the Drude conductivity $\sigma$ 
arising from the proximity effect. This approximation is applicable to 
the second order in the $NS$ interface transparency.

\begin{figure}[t]
\includegraphics[width=0.9\columnwidth]{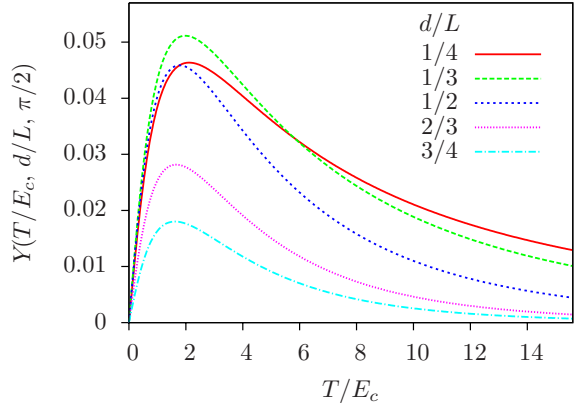}
\caption{(Color online)
The function  $Y(T/E_c,d/L,\pi/2)$ from Eq.~(\ref{Sps}) versus the ratio $T/E_c$
for the parallelogram interferometer with the symmetric arms $\ell_1=\ell_2$. 
The thermopower is maximal for $T\sim 2E_c$, where $E_c=D/L^2$
is the Thouless energy associated with the normal-metal wire $N$.
At high temperatures the thermopower decays according to the stretched 
exponential law (\ref{Slarge}) that is determined by the Thouless energy 
$E_c'=D/d^2$ associated with the distance between the $NS$ junctions.
}
\label{fig:Yplot}
\end{figure}

In order to calculate $J_0$ in Eq.~(\ref{J0final})
we substitute $\mu$ found from Eq.~(\ref{s1}) into Eq.~(\ref{J0final}). 
Disregarding the terms that are proportional to $h_{\mu T}$ we 
obtain the equation on the imbalance $\mu$ in the following form
\beq
&&\sin\varphi\,\frac{d}{L} \int  d\ep\, h_T \im F_{12} \n\\
&&\quad
=\int d\ep\, h_\mu\re\lt(\frac{1}{2}(F_1+ F_2)+F_{12}\cos\varphi\rt).
\eq
Expanding this equation to the first order in $\mu$ and the temperature gradient 
we obtain
\be
\label{mufinal}
\mu=A (T_1-T_2) \sin\varphi, 
\e
where $T=(T_1+T_2)/2$ is the mean temperature in the system and
the coefficient $A$ is given by
\be
\label{A}
A=\frac{(d/L)\,\int d\ep\, \ep \cosh^{-2}\frac{\ep}{2T}\, \im F_{12}}{
T\int d\ep\, \cosh^{-2}\frac{\ep}{2T}\, \re (F_1+F_2+2F_{12} \cos\varphi)}.
\e
We note that the coefficient $A$ is finite even for vanishing interface 
transparency parameters $\alpha_{1,2}$.
From Eqs.~(\ref{J0final},\ref{mufinal}) we find the thermopower
\be
S=\frac{A d\, \sin\varphi}{8 e T}
\int d\ep\,\cosh^{-2}\frac{\ep}{2T}\, \re (F_2-F_1),
\label{Sfinal}
\e
that behaves roughly as $\sin{\varphi}$. The thermopower is finite 
only if there is a difference between $F_1$ and $F_2$, hence the device asymmetry
is required. The Equation~(\ref{Sfinal}) describes the analog 
of the Pethick-Smith effect in a superconducting proximity system.  

The imbalance given by Eq.~(\ref{mufinal}) is a non-monotonous 
function of the mean temperature $T$ that reaches its maximal value 
at $T\sim E_c=D/L^2$ and decays as a stretched exponent for $T\gg E'_c=D/d^2$.
It is worth noting that the imbalance itself is finite even for a 
left-right symmetric setup and in the limit of small transparency 
of the $NS$ interfaces. The temperature dependence of $\mu$ 
is determined by that of the coefficient $A$ in Eq.~(\ref{A}).
The energy integrals in Eq.~(\ref{A}) can be rewritten as sums over
the Matsubara frequencies. This greatly simplifies the calculation of  
the asymptotic behavior of $A$ at large temperatures. 

Let us estimate the coefficient $A$ for a symmetric setup with $\ell_1=\ell_2=\ell$ and $\alpha_1=\alpha_2$.
In the low temperature limit $T\ll E_c$ we obtain 
\be
\label{Asmall}
A=c_\varphi\frac{T\, L\ell}{D},\quad c_\varphi=\frac{\pi^2}{18}\frac{1-2(\ell/L)^2}{1-(2\ell/L)^2\sin^2(\varphi/2)}.
\e
For high temperatures $T\gg E_c'$ we find from the Matsubara representation 
of Eq.~(\ref{A}) that
\be
\label{Alarge}
A=c_1\frac{d}{\ell}\lt(\sqrt{2\pi T/E_c'}-1\rt)e^{-\sqrt{2\pi T/E_c'}},
\e
where
\be
c_1=\frac{\pi}{(2-2^{-1/2})\zeta(3/2)}\approx 0.93.
\e
The analysis shows that the coefficient $A$ reaches 
its maximal value at $T\simeq E_c=D/L^2$. 

For  $T\gg E_c'$ the energy integral in Eq.~(\ref{Sfinal}) is decaying as $T^{-1/2}$, 
provided $\alpha_1\neq \alpha_2$. This decay crosses over to a stretched exponential 
decay for $\alpha_1=\alpha_2$ if the asymmetry is caused merely 
by a difference between $\ell_1$ and $\ell_2$. Therefore an asymmetry in 
the transparency parameter is of a greater importance to the thermopower than that
in the arm lengths. For an equal arm interferometer $\ell_1=\ell_2=\ell$ with 
$\alpha_1\neq \alpha_2$ we obtain from 
Eqs.~(\ref{Asmall},\ref{Alarge}) in the limit $T\ll E_c=D/L^2$ that
\be
S=\frac{d\, c_\varphi (\alpha_2^2-\alpha_1^2)\sin\varphi}{4 e}\lt(1+\frac{d}{L}\rt) \frac{T\ell^2}{D}.
\e
For high temperatures, $T\gg E_c' =D/d^2$, we get 
\be
\label{Slarge}
S=\frac{\pi d(\alpha_2^2-\alpha_1^2)\sin\varphi}{8 e}\frac{d^2}{L^2}\lt(1-\sqrt{\frac{E_c'}{2\pi T}}\rt)
e^{-\sqrt{2\pi T/E_c'}}.
\e
The stretched exponential decay of the Pethick-Smith thermopower 
at high temperatures is entirely due to the behavior of the imbalance $\mu$. 

For the interferometer with $\ell_1=\ell_2=\ell$ it is convenient to introduce the 
dimensionless function $Y$ by the following relation
\be
\label{Sps}
S=\frac{\pi d (\alpha_2^2-\alpha_1^2)\sin\phi}{8 e}\, Y\lt(\frac{T}{E_c},\frac{d}{L},\varphi \rt).
\e
We see from Eqs.~(\ref{Ssmall},\ref{Slarge}) that $Y$ is a very smooth function of 
the order parameter phase difference $\varphi$. The dependence on $\varphi$
disappears in the limit $T\gg E_c'$. We plot the temperature 
dependence of $Y$ in Fig.~\ref{fig:Yplot} for $\varphi=\pi/2$ and $d/L=1/4,1/3,1/2,2/3,3/4$. 
The thermopower  reaches its maximal value in the interferometer with 
$\ell_1=\ell_2=d=L/3$ at the temperature $T\simeq 2 E_c$.
Our results are consistent with the experiment of Ref.~\onlinecite{Eom1998} 
and with the numerical analysis of Refs.~\onlinecite{Virtanen2004a,Virtanen2004b,VP2005}. 
The magnetic flux dependence of the thermopower is always antisymmetric and
is essentially given by  $\sin 2\pi \Phi/\Phi_0$.  A small deviation from this law 
may arise at low temperatures $T\lesssim E_c'$. The strength of the proximity induced 
thermoelectric effect in the parallelogram interferometer is restricted by the 
asymmetry parameter  $\alpha_1^2-\alpha_2^2$ and by the maximal value of 
the dimensionless function $Y$ that is approximated by $Y_{\rm max}\approx 0.05$. 

\section{thermoelectric effect in the house interferometer}
\label{sec:house}

\begin{figure}[t]
\includegraphics[width=0.9\columnwidth]{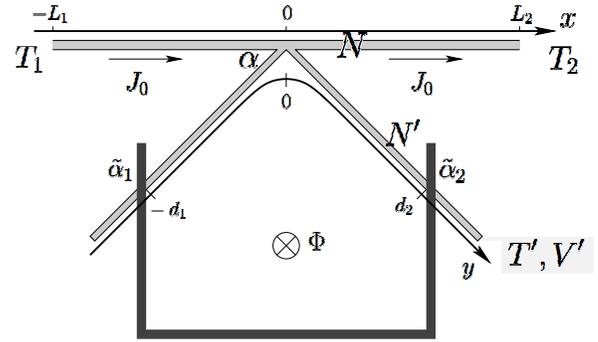}
\caption{
The house interferometer. The wires $N$ and $N'$ are normal metal wires.
The dark wire is superconducting. The transparency parameters $\alpha$,
$\tilde{\alpha}_{1,2}$ are defined by the ratio between the transmission probability 
per channel and the effective barrier length. Both symmetric and antisymmetric 
dependence of the thermopower on the magnetic flux $\Phi$ is possible in this setup.
The antisymmetric effect, however, requires $d_1\neq d_2$. A charge imbalance
between the superconducting wire and the temperature reservoirs
is necessary for the symmetric thermopower. 
}
\label{fig:house}
\end{figure}

In this Section we apply the same theory to the house interferometer depicted in
Fig.~\ref{fig:house}. In our description of the system we introduce 
two normal-metal wires, $N$ and $N'$, that are connected 
via a single tunnel junction with the transparency parameter $\alpha$. 
In the actual experiment of Ref.~\onlinecite{Eom1998} the junction $NN'$ is ballistic
because the metallic wires are produced in a single lithography circle. 
We, however, expect that the symmetry of the thermoelectric coefficient
is not affected by the detailed characteristics of the $NN'$ junction. 

Different space variables $x$ and $y$ are introduced as the coordinates 
along the wires $N$ and $N'$, correspondingly. The proximity effect 
is described by the Green's functions $\check{g}_c(y)$ in $N'$ and 
$\check{g}(x)$ in $N$.  The $NN'$ junction is located at $x=y=0$ and 
the wire $N'$ is contacted to the superconductor 
at $y=-d_1$ and $y=d_2$. The transparency 
parameters $\tilde{\alpha}_1$ and $\tilde{\alpha}_2$ of the $N'S$
interfaces  are introduced in the same way as in the previous Section. 

The thermopower measurements of Ref.~\onlinecite{Eom1998}
are shown in Fig.~\ref{fig:exp}b.  In the experiment the 
distance between the $N'S$ is given by
$d=d_1+d_2\sim 2\mu m$ that is smaller than the phase coherence 
length $L_\phi \sim 4.6\mu m$ at $T=38mK$. We, therefore, expect 
that the setup is adequately described by the Usadel equation. 

The charge conservation condition in the loop formed by the 
superconductor and the $N'$ wire is expressed as
\be
\label{conserv}
\delta\Big[j^e\Big]_0=0.
\e
Thus, there exists no Pethick-Smith like contribution to $J_0$ 
that is analogous to the first term in Eq.~(\ref{s3}).
Another words the mechanism of the superconductor charging 
due to a supercurrent flow is absent in the house interferometer. 
Unlike in the parallelogram interferometer, the other two contributions 
to $J_0$ in Eq.~(\ref{s3}) survive, 
\be
\label{J0house}
J_0=\frac{\sigma}{8e}\int d\ep
\Bigg\{\frac{h_{T}}{2L}\frac{1}{L}\il_{-L_1}^{L_2}
dx\,Y -\frac{h_{\mu T}}{2L}\frac{1}{L}\il_{-L_1}^{L_2}
dx\,M_+\Bigg\}.
\e
It is worth noting that the expression (\ref{J0house}) for the dissipative current between the
temperature reservoirs is valid for any interface transparency. 

The first term of Eq.~(\ref{J0house}), which is proportional to $h_{T}$, is antisymmetric 
in the magnetic field and is further referred to as the interference contribution. 
This contribution is insensitive to the imbalance and vanishes in the parallelogram interferometer 
because of the property (\ref{zero}). It remains, however, finite in the house interferometer 
where the length of the interfering trajectories is fixed by the distances $d_1$ 
and $d_2$ between the $N'S$ interfaces and the $NN'$ junction. 

\begin{figure}[t]
\includegraphics[width=0.8\columnwidth]{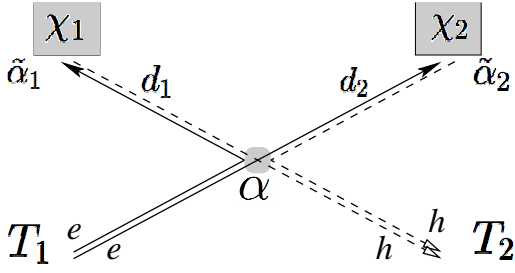}
\caption{
An example of interfering trajectories that contribute to the thermoelectric 
response of the house interferometer.  
}
\label{fig:track}
\end{figure}

Two interfering quasiparticle trajectories, which contribute to the heat transfer between the 
reservoirs, are depicted schematically in Fig.~\ref{fig:track}. The scattering amplitude 
of the process can be written as
\be
{\cal A}_\ep\sim \alpha^2 |\tilde{\alpha}_1 e^{i(k_e-k_h)d_1+i\chi_1} 
+ \tilde{\alpha}_2e^{i(k_e-k_h)d_2+i\chi_2} |^2,
\e
where $k_{e,h}=k_F\pm \ep/v_F$ are the wave vectors of electrons and holes. Here we 
took into account that the Andreev reflection process is of a second order in the 
$N'S$ interface transparency. The corresponding contribution to the thermopower 
is proportional to the energy derivative of the scattering amplitude 
${\cal A}_\ep$ at the Fermi energy,
\be
\label{estimate}
S\sim \lt.\frac{d{\cal A}_\ep}{d\ep}\rt|_{\ep=0}\propto \alpha^2\tilde{\alpha}_1\tilde{\alpha}_2(d_2-d_1) \sin\varphi,
\e
which is a manifestly antisymmetric function of the magnetic flux.  
The result of Eq.~(\ref{estimate}) is finite if $d_1\neq d_2$, while the difference 
in the transmission parameters $\tilde{\alpha}_1$ and $\tilde{\alpha}_2$ 
is neither necessary nor sufficient for the observation of the interference effect. 
Integration over all possible trajectories in the parallelogram interferometer
lead to the complete suppression of interference that is reflected in Eq.~(\ref{zero}).

The interference contribution to the thermopower discussed above has not been observed 
in experiment, because it is overwhelmed by the second term in Eq.~(\ref{J0house}) which 
we refer to as the symmetric contribution. This term is proportional to $h_{\mu T}$ 
and symmetric with respect to the magnetic flux provided $\mu$ is flux independent. 
It remains to be finite for a left-right symmetric device 
but vanishes for $\mu=0$.  

The experimental measurements of Ref.~\onlinecite{Eom1998} shown in Fig.~\ref{fig:exp}b
can be regarded as demonstration of a finite imbalance state formed in the house interferometer. 
The experiment suggests that a constant imbalance $\mu$ is largely independent
on the temperature gradient in the wire $N$ as well as on the supercurrent in the loop.
Since $\mu=0$ is the only solution to the charge conservation condition (\ref{conserv}), 
the origin of such a steady state is not entirely clear. 
We stress, however, that a tiny charge imbalance can originate in the usual 
thermoelectric effect provided $N'$ wire is in an equilibrium state
with a temperature $T'$ that is smaller than both $T_1$ and $T_2$ due to the phonon cooling.
In this case we can estimate $\mu=e(T-T')S_{N'}$, where $S_{N'}$ is the small 
thermopower of the $N'$ wire. The tiny charge imbalance $\mu$ will be seen to have 
a great effect on the thermopower due the proximity effect enhancement factor. In what follows 
we simply keep $\mu$ as a phenomenological parameter of the quasiclassical theory. 
 
We start by calculating the anomalous component $f^R(x)$ of the Green's function 
in the normal metal wire $N$. We again take advantage of 
the linearized Usadel equation (\ref{UsadelfR}) 
and employ the rigid boundary condition
\be
\label{cond_house}
\delta\lt[\frac{df^R}{dx}\rt]_0=-\alpha f_{c0}^R,
\e
where $f_{c0}^R=f_c^R(y=0)$ is the value of the anomalous Green's function 
at the junction.
  
In addition to Eq.~(\ref{cond_house}) we use  Eq.~(\ref{fra_zero}) 
and the continuity of $f^R(x)$ at $x=0$. As the result we obtain
\be
\label{fRhouse}
f^R=\frac{\alpha f_{c0}^R}{z \sinh{zL}}\! \lt\{
\ba{cr} 
\sinh{z L_2}\sinh{z(L_1\!+\!x)}, & -L_1<x<0\\
\sinh{z L_1}\sinh{z(L_2\!-\!x)}, & \ \ 0<x<L_2\\
\ea
\rt..
\e 
We calculate the  spatial integrals in  Eq.~(\ref{J0house})
with the help of Eq.~(\ref{fRhouse}) as
\beml
\beq
&&\frac{1}{L}\int_{-L_1}^{L_2}\!\!\!dx\,
(f^R\bar{f}^A\pm f^A\bar{f}^R)=\frac{D}{2\ep}(f_{c0}^R\bar{f}_{c0}^A\pm
f_{c0}^A\bar{f}_{c0}^R)
\im F, \n \\ 
&& F=\alpha^2 \frac{\sinh zL_1 \sinh zL_2}{z L \sinh z L}, \qquad L=L_1+L_2.
\eq
\eml
It is worth noting that the first term of Eq.~(\ref{J0house}) is not necessarily vanishing 
unlike in the parallelogram interferometer. 

We evaluate the second term of Eq.~(\ref{J0house}) 
to the leading order in the $N'S$ interface  transparency   
by calculating the following integral
\beq
&&\frac{1}{L}\int_{-L_1}^{L_2}\!\!\!dx\,
(f^R\bar{f}^R+ f^A\bar{f}^A)=\frac{D}{2\ep}\im f_{c0}^R\bar{f}_{c0}^R (K-F),\n\\
&& K=\alpha^2\frac{L_1 \sinh^2 zL_2 +L_2\sinh^2 zL_1}{L \sinh^2 zL},
\eq
that is also expressed through the the anomalous Green's function in the $N'$ wire.
Thus, from Eq.~(\ref{J0house}) we find, to the second order in $f_{c0}$, that
\beq
\label{final}
&&J_0=\frac{\sigma}{8e}\int d\ep\, \frac{D}{2\ep} \Big\{ 
\frac{h_{T}}{2L} (f_{c0}^R\bar{f}_{c0}^A-f_{c0}^A\bar{f}_{c0}^R)\im F \\ \n
&&
-\frac{h_{\mu T}}{2L} 
\lt((f_{c0}^R\bar{f}_{c0}^A+f_{c0}^A\bar{f}_{c0}^R)\im F
+\im f_{c0}^R\bar{f}_{c0}^R (F-K)\rt) \Big\}.
\eq
It remains to determine the function $f_c(y)$ from the Usadel equation in the wire $N'$.
Since we calculate the current $J_0$ to the leading order in the interface transparency parameters   
the tunneling between $N$ and $N'$ can be disregarded in the Usadel equation. 
Taking advantage of the rigid boundary conditions in the form
\be
\lt.\frac{d f_c^R}{d y}\rt|_{-d_1}=-\tilde{\alpha}_1 e^{i \chi_1},\qquad
\lt.\frac{d f_c^R}{d y}\rt|_{d_2}=\tilde{\alpha}_2 e^{i \chi_2}, 
\e
we obtain
\be
f^R_c(y)=\tilde{\alpha}_1 e^{i\chi_1} \frac{\cosh z(d_2-y)}{z \sinh z d}
+\tilde{\alpha}_2  e^{i\chi_2} \frac{\cosh z(d_1+y)}{z \sinh z d},
\e
where $d=d_1+d_2$. Thus, the value of the anomalous Green's function at the contact
is given by
\be
\label{crude}
f^R_{c0}=\tilde{\alpha}_1 e^{i\chi_1} \frac{\cosh z d_2}{z \sinh z d}+
\tilde{\alpha}_2 e^{i\chi_2} \frac{\cosh z d_1}{z \sinh z d}.
\e
Substitution of Eq.~(\ref{crude}) to Eq.~(\ref{final}) completes the calculation of the
thermoelectric effect. 

Let us now analyze two different terms in Eq.~(\ref{final}) separately.
We first let  $\mu=0$ so that $h_{\mu T}=0$ and the only contribution to Eq.~(\ref{final})
is antisymmetric in the magnetic flux. From Eq.~(\ref{crude})
we find  the expression
\be
\label{interference}
f_{c0}^R\bar{f}_{c0}^A-f_{c0}^A\bar{f}_{c0}^R=
4\tilde{\alpha}_1\tilde{\alpha}_2 \sin\varphi \frac{\im \lt(\cosh z d_1 \cosh z^* d_2\rt)}{|z \sinh zd|^2}.
\e
In order to obtain the thermopower $S=(J_0/\sigma)(T_2-T_1)^{-1}$
we substitute Eq.~(\ref{interference}) to Eq.~(\ref{final}) and expand $h_T$
to the linear order in the temperature gradient as
\be
h_T = -\frac{\ep (T_1-T_2)}{T^2\cosh^2\frac{\ep}{2T}}.
\e
The resulting expression for $S$ reads
\be
\label{Sinter}
S=\frac{\alpha^2\tilde{\alpha}_1\tilde{\alpha}_2 \sin\varphi \, D d^2}{8 e L}\frac{B}{T}
\e
with the coefficient $B$ given by
\be
\label{B}
B=\int d\ep 
\frac{\im\lt(\cosh zd_1\,\cosh z^* d_2\rt)}{|z d \sinh z d|^2\, T\cosh^2\frac{\ep}{2T} }
\im\frac{\tanh(zL/2)}{zL\sinh{zL}},
\e
where we let $L_1=L_2=L/2$ for simplicity. 
The Equation~(\ref{Sinter}) describes the interference contribution to the thermopower. 
The result of Eq.~(\ref{Sinter}) is indeed proportional to $\alpha^2\tilde{\alpha}_1 \tilde{\alpha}_2\sin\varphi$ 
and is vanishing for $d_1=d_2$ in accordance with the estimate (\ref{estimate}).
Moreover, in the limit $T\ll {\rm min}\,\lt\{E_c,E_c' \rt\}$, where $E_c=D/L^2$ and
$E_c'=D/d^2$, we find
\be
\label{Ssmall}
S=\frac{\alpha^2\tilde{\alpha}_1\tilde{\alpha}_2(d_2-d_1)D L}{198 e T d}\sin\varphi,
\e 
that is equivalent to Eq.~(\ref{estimate}).  We note that the divergence 
of Eq.~(\ref{Ssmall}) in the limit $T\to 0$ is regularized by the high order processes 
in the parameter $\alpha$. The result of Eq.~(\ref{Ssmall}) is, therefore, valid 
only for $T\gtrsim \alpha^2 D$. More accurate expression 
can be obtained by solving the non-linear Usadel equation in the normal-metal wire $N'$.

For higher temperatures $T\gg {\rm max}\,\lt\{E_c,E_c' \rt\}$ it is legitimate
to substitute $\cosh (\ep/2T)\to 1$ in Eq.~(\ref{B}). 
The remaining integral gives the dependence of 
the thermopower on the distances $L$, $d_1$, and $d_2$.
Thus, unlike the Pethick-Smith contribution (\ref{Sps}) to the thermopower, 
the interference contribution has a  monotonous temperature dependence
and decays as the power law $T^{-2}$  at high temperatures. It is also the
only contribution that remains finite for $\mu=0$. 

Let us consider now the symmetric contribution to Eq.~(\ref{final}) that emerges for a 
finite $\mu$. Since this contribution does not rely on the device asymmetry, 
we let for simplicity $d_1=d_2=d/2$. Following the discussion above we 
treat the imbalance $\mu$ as a constant in our quasiclassical analysis 
and calculate the thermopower with the assumption that $\mu$ is independent on
the temperature gradient $T_2-T_1$. The substitution of Eq.~(\ref{crude}) 
to Eq.~(\ref{final}) gives
\be
\label{final2}
J_0=-\frac{\sigma}{4e}\frac{D(\tilde{\alpha}_1^2+\tilde{\alpha}_2^2
+2\tilde{\alpha}_1\tilde{\alpha}_2\cos\varphi)}{TL}\int d\ep\,h_{\mu T} P,
\e
where
\be
\label{diverge}
P=\frac{T}{16 \ep}\im\lt\{\frac{F}{|z\sinh zd/2|^2}+\frac{F-K}{2(z\sinh zd/2)^2} 
\rt\}.
\e
Thus, for the completely symmetric case $\tilde{\alpha}_1=\tilde{\alpha}_2$ 
the thermopower has an overall phase dependent factor $1+\cos\varphi$,
that qualitatively agrees with the experimental curve shown in Fig.~\ref{fig:exp}b. 
In general, the minimal absolute value of the symmetric contribution to the
thermopower is determined by the asymmetry parameter 
$(\tilde{\alpha}_1-\tilde{\alpha_2})^2$.   

\begin{figure}[t]
\includegraphics[width=0.9\columnwidth]{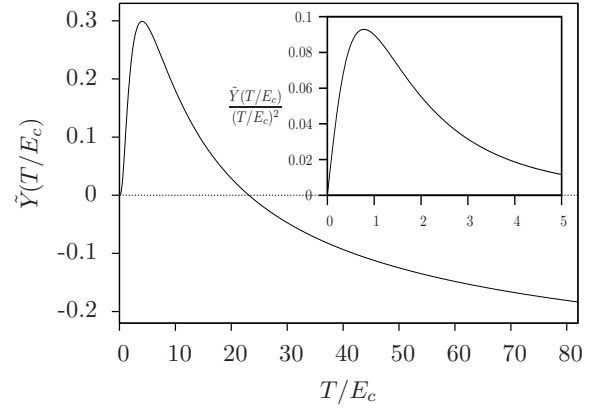}
\caption{
The function  $\tilde{Y}(T/E_c)$ from Eqs.~(\ref{Shouse},\ref{Y}) 
versus the ratio $T/E_c$ for the house interferometer with $d_1=d_2$. 
The function is maximal for $T\approx 4 E_c$ and changes sign at 
$T\approx 24 E_c$, where $E_c=D/L^2$ is the Thouless energy 
associated with the normal-metal wire $N$.
The inset shows the temperature dependence of the thermopower 
in Eq.~(\ref{Shouse}) under the assumption that the imbalance $\mu$ 
and the length $L$ are temperature independent.
The thermopower is maximal at $T\approx E_c$ and decays as 
$T^{-2}$ for $T \gg 24 E_c$. 
}
\label{fig:Hs}
\end{figure}

We note, however, that the integral in Eq.~(\ref{diverge}) diverges at small energies.
This divergence is regularized by the high order processes in the parameter $\alpha$.
In order to get a well-defined expression we need to solve the non-linear Usadel equation 
in $N'$ that is a daunting task. We give instead an analytical estimate to the 
symmetric thermopower by replacing the $N'$ wire with a diffusive quantum dot. 
In this model the Green's function, $\check{g}_c$, in the dot can be obtained 
in the spirit of the Nazarov's circuit theory\cite{Nazarov1994} 
from the matrix current conservation condition
\be
[\tilde{\alpha}_1\hat{g}_{S,1}^R+
\tilde{\alpha}_1\hat{g}_{S,2}^R+
\alpha\sigma_z, g_c^R]=0.
\e
Resolving the condition for $\hat{g}_c^R$ we obtain
\be
\label{fcRnazarov}
f_{c}^R=(\tilde{\alpha}_1e^{i\chi_1}+\tilde{\alpha}_2e^{i\chi_2})u^{-1},
\e
with 
\beq
\n
u&=&\lt((\tilde{\alpha}_1+\tilde{\alpha}_2)(-i\ep/\Delta)+
\alpha\sqrt{1+(-i\ep/\Delta)^2}\rt)^2\\
&+& \tilde{\alpha}_1^2+\tilde{\alpha}_2^2
+2\tilde{\alpha}_1\tilde{\alpha}_2\cos\varphi.
\label{uu}
\eq
In this approximation the anomalous Green's function 
$f_{c}^R=f_{c0}^R$ acquires no spacial dependence on $y$. 
The function  $f_{c}^R$  remains to be small in the parameters $\tilde{\alpha}_1$ and $\tilde{\alpha}_2$,
therefore Eq.~(\ref{final}) is still a legitimate approximation for the thermopower
to the leading order in $\tilde{\alpha}_{1,2}$. 
Substituting $f_{c0}^R=f_c^R$ from Eq.~(\ref{fcRnazarov}) to Eq.~(\ref{final}) 
we arrive at the result (\ref{final2}) with
\be
\label{Pfinite}
P=\frac{T}{\ep}\im\lt\{\frac{F}{|u|}+\frac{F-K}{2 u} \rt\}.
\e
In order to be consistent we have to take the limit $\ep\ll \Delta$ and 
$\tilde{\alpha}_a\ll \alpha$ in the expression (\ref{uu}), hence
\be
u=\alpha^2,
\e
and the parameter $\alpha$ cancels out completely in Eq.~(\ref{Pfinite}).
In the limit $\mu\ll T$ and $T_1-T_2\ll T$ we expand $h_{\mu T}$ as
\be
\label{exp}
h_{\mu T}\simeq \frac{\mu(T_1-T_2)}{T^2 \cosh^2\ep/2T}\lt(1-\frac{\ep}{T}\tanh\frac{\ep}{2T}\rt)
\e
and substitute this expression into Eq.~(\ref{final2}) with the function $P$ given by Eq.~(\ref{Pfinite}). 
This provides us with the final result for the thermopower $S=(J_0/\sigma)(T_2-T_1)^{-1}$
of the house interferomter
\be
\label{Shouse}
S=-\frac{D\mu(\tilde{\alpha}_1^2+\tilde{\alpha}_2^2
+2\tilde{\alpha}_1\tilde{\alpha}_2\cos\varphi)}{4eT^2 L}\, \tilde{Y}\lt(\frac{T}{E_c}\rt),
\e
where $E_c= D/L^2$ and the dimensionless function $\tilde{Y}(T/E_c)$ 
is given by 
\be
\label{Y}
\tilde{Y}=\il_{-\infty}^{\infty} dx\, \frac{2x\tanh x -1}{4x\cosh^2x}
\im\lt\{\frac{3\tanh\kappa_x}{2\kappa_x}-\frac{1}{\cosh^2\kappa_x}\rt\},
\e
with $\kappa_x\equiv \sqrt{-ix T /E_c}$. In the limit $T\ll E_c$ we find 
$\tilde{Y}(T/E_c)=(187 \pi^2/7560) (T/E_c)^3$. The function $\tilde{Y}$ determines the temperature
dependence of the symmetric contribution to the thermopower (see Fig.~\ref{fig:Hs})
provided $\mu$ is a temperature independent constant. This contribution is not suppressed 
by any asymmetry factor unlike the Pethick-Smith thermopower (\ref{Sps}) 
and the interference contribution (\ref{Sinter}).
Moreover, the symmetric contribution is neither monotonous nor sign-definite function of temperature. 
At $T\sim E_c$ the thermopower approaches its maximum value, which can be estimated as
\be
\lt.S\rt|_{T\simeq E_c} \approx -0.1 \frac{\mu L^3}{4eD}(\tilde{\alpha}_1^2+\tilde{\alpha}_2^2
+2\tilde{\alpha}_1\tilde{\alpha}_2\cos\varphi).
\e
Thus, the presence of a very small imbalance $\mu$ leads to large observable 
magnetic flux dependence of the thermopower. The absolute value of 
the thermopower at $T\sim E_c$ and $\Phi=0$ is determined by the
parameter $\mu L/(\ell_B E_c)$, where $\ell_B$ is the effective barrier 
length that is orders of magnitude smaller than $L$. This is the proximity effect
enhancement factor that gives rise to an anomalously strong sensitivity 
of the thermopower to the charge imbalance $\mu$.

We also see from Eqs.~(\ref{Shouse},\ref{Y}) that the thermopower 
changes sign at $T\approx 24 E_c$ and decays as
$T^{-2}$ at higher temperatures since $\tilde{Y} \to -2/5 $ for $T\gg 24 E_c$.
For higher temperatures the length $L$ has to be substituted 
by the phase coherence length $L_\phi$ 
provided the latter is smaller than the distance between the reservoirs.
The temperature dependence of $L_\phi$ complicates the direct comparison 
to the experimental data. In particular this dependence should strongly enhance 
the reversed thermopower for $T\gtrsim 24 E_c$. 
Nevertheless, the sign reversal behavior of the the symmetric thermopower 
observed by Parsons {\it et al.}\cite{Parsons2003,Parsons2003b} is 
in a qualitative agreement with the result of Eq.~(\ref{Shouse}). 

A thorough experimental test of the presented theory can be performed 
with the help of an independent experimental control 
over the imbalance $\mu$ that has a strong effect on the sign and the magnitude 
of the symmetric contribution to the thermopower in the house interferometer.
For the parallelogram interferometer
the symmetric contribution is described by the last term in Eq.~(\ref{s3}),
that is of the second order in the temperature gradient. This term can be 
roughly estimated by substituting $\mu$ from Eq.~(\ref{mufinal}) to Eq.~(\ref{Shouse}). 
This second order effect reverses sign with increasing temperature and is antisymmetric 
in $\Phi$ due to the magnetic flux dependence of the imbalance in the parallelogram 
interferometer. It is likely that the effect of this type has been observed 
in the experiments by Parsons {\it et al.}\cite{Parsons2003,Parsons2003b}

\section{conclusion}
\label{sec:conclusion}

In this paper we applied quasiclassical theory to study the origin of the 
thermoelectric effects in Andreev interferometers.
The theory predicts three additive contributions 
to the electric current expressed by Eqs.~(\ref{Je},\ref{current_e}) 
in the presence of the superconducting proximity effect. 
Depending on the geometry of Andreev interferometer and its parameters 
any of these contributions may dominate the quasiclassical 
thermoelectric response of the normal-metal wire.
One can classify the observed thermoelectric effect 
by its dependence on the temperature and the magnetic flux piercing the interferometer.
The last term in Eq.~(\ref{current_e}), which describes the proximity induced supercurrent, 
and the second term, which is related to the interference contribution, are antisymmetric in the magnetic flux. 
Even though the supercurrent cannot flow between the normal metal reservoirs 
the last term in Eq.~(\ref{current_e}) can, nevertheless, contribute to the thermopower
by means of the proximity-induced Pethick-Smith effect. This effect takes place provided 
the temperature gradient is aligned in a part of the normal metal wire 
with the supercurrent.  This situation is realized in the parallelogram interferometer 
considered in the Section~\ref{sec:para}. 

The left-right asymmetry of 
the parallelogram interferometer is responsible for a difference in the dissipative 
charging currents flowing from the normal-metal reservoirs to the superconductor. 
As the result of this asymmetry a compensating dissipative current $J_0$ (\ref{J0final})
is flowing between the normal-metal reservoirs.
The analytical expression for the thermopower in the case of weak proximity effect
is given by Eqs.~(\ref{A},\ref{Sfinal}). The proximity-induced Pethick-Smith 
effect demonstrates a 
non-monotonous temperature dependence with a maximum at $T\simeq 2 E_c$ 
(see Fig.~\ref{fig:Yplot}) and is characterized by the stretched exponential decay 
for high temperatures (\ref{Slarge}). The effect does not rely on the quasiparticle 
phase coherence in the normal-metal wire. 
The most important ingredient of the theory is the difference in the electron distribution 
function near the first and the second $NS$ interface in the parallelogram interferometer. 

In contrast, the interference contribution to the thermopower, which
is given by the second term in Eq.~(\ref{current_e}), does require a phase coherence 
in the normal-metal wire. This contribution is the only one that exists in the absence 
of a charge imbalance between the superconductor and the normal metal. 
This effect cannot be seen in the parallelogram interferometer but might be observed in 
the house one provided $d_1\neq d_2$ in Fig.~\ref{fig:house}. The interference 
contribution  is characterized by a monotonous temperature decay, that is
estimated in Eqs.~(\ref{Sinter},\ref{B}) in the Section~\ref{sec:house}.  

Finally, the first term in Eq.~(\ref{current_e}) is regarded as the symmetric contribution.
It is largely insensitive to the left-right symmetry of the device  and gives rise 
to the thermopower that is symmetric with respect to the magnetic flux 
(with the assumption that $\mu$ is flux independent). 
In the Section~\ref{sec:house} we argue that the existing experiments 
indicate the presence of such a steady state imbalance in the house interferometer.
The symmetric contribution is characterized by a peculiar temperature 
dependence in Eqs.~(\ref{Shouse},\ref{Y})
that is neither monotonous nor sign-definite. This contribution strongly affects 
the thermoelectric response of the parallelogram interferometer 
in the second order with respect to the temperature gradient.

\begin{acknowledgements}
The author acknowledges  discussions with W.~Belzig, A.~A.~Golubov, and Yu.~V.~Nazarov
and the kind hospitality of the Isaac Newton Institute for Mathematical Sciences 
of the Cambridge University, where this work has been partially written. 
The author thanks Venkat Chandrasekhar for reading the manuscript prior publication
and is especially grateful to Igor Aleiner for numerous discussions and
motivation of the present study.
\end{acknowledgements}

\end{document}